\documentclass{emulateapj}

\shorttitle{The Smith Cloud}
\shortauthors{Nichols \& Bland-Hawthorn}

\newcommand{\sol}{\odot}
\newcommand{\HI}{H{\small I}}
\newcommand{\kms}{\ km s$^{-1}$}

\begin{document}
\title{The Smith Cloud: high-velocity accretion and dark-matter confinement}
\author{M. Nichols and J. Bland-Hawthorn}
\affil{Sydney Institute for Astronomy, School of Physics, The University of Sydney, NSW 2006, Australia}
\email{m.nichols@physics.usyd.edu.au}

\begin{abstract}
The Smith Cloud is a massive system of metal-poor neutral and ionized gas (M$_{\rm gas}$ $\ga 2\times 10^6$~M$_\odot$) that is presently moving at high velocity (V$_{\rm GSR}$ $\approx$~300~km~s$^{-1}$) with respect to the Galaxy at a distance of 12~kpc from the Sun.
The kinematics of the cloud's cometary tail indicates that the gas is in the process of accretion onto the Galaxy, as first discussed by \citet{Lockman08}.
Here, we re-investigate the cloud's orbit by considering the possibility that the cloud is confined by a dark matter halo.
This is required for the cloud to survive its passage through the Galactic corona.
We consider three possible models for the dark matter halo (NFW, Einasto, Burkert) including the effects of tidal disruption and ram-pressure stripping during the cloud's infall onto and passage through the Galactic disk.
For the NFW and Einasto \footnote{The Einasto profile is sometimes referred to as the Sersic profile 
\citep[e.g.][]{Zait08}. These have the same functional form but since 
the Sersic profile arises from a projected distribution, we prefer to 
use a radial function. We note, however, that the
projection of the Einasto function is nontrivial analytically 
\citep{Mazure02}.} dark-matter models, we are able to determine reasonable initial conditions for the Smith Cloud, although this is only marginally possible with the Burkert model.
For all three models, the progenitor had an initial (gas+dark matter) mass that was an order of magnitude higher than inferred today.
In agreement with \citeauthor{Lockman08}, the cloud appears to have punched through the disk $\approx$~70~Myr ago. 
For our most successful models, the baryon to dark matter ratio is fairly constant during an orbital period but drops by a factor of $2-5$ after 
transiting the disk. The cloud appears to have only marginally survived its transit, and is unlikely to retain its integrity during the next transit $\approx$~30~Myr from now.
\end{abstract}

\section{Introduction}
The Smith Cloud is perhaps the best known and most studied member of the high-velocity clouds (HVCs).
While covering over a third of the sky, little is known about the nature of these clouds, with many suggestions as to their origin.
While we now know the distances to a number of HVCs, their orbits are more difficult to determine.
Since its discovery in \citeyear{Smith63}, the Smith Cloud was originally believed to be an extension of the Galactic disk \citep{Smith63} but it was later shown to be separated kinematically from the Galactic disk \citep{Lockman84} leading to its subsequent classification as an HVC \citep{Wakker97}.

While the origin of the cloud is still unknown, with suggestions of gas belonging to an infalling dwarf galaxy \citep{Bland-Hawthorn98} or gas lifted from the Galactic disk \citep{Sofue04}, the distance, mass and velocity of the cloud are now well established.
Distance indicators  based on stellar absorption-line bracketing, H$\alpha$ flux, and the kinematic distance of gas interacting with the cloud, provide a consistent estimate of $12.4\pm1.3$~kpc \citep{Lockman08, Putman03, Wakker08}.
The H{\small{}I} mass of the cloud is found to be $1\times 10^6$~M$_\sol$ \citep{Lockman08}, with WHAM data indicating a comparable or greater amount in ionized hydrogen \citep{Hill09}. The velocity components have also been determined \citep{Lockman08} allowing the orbit of the cloud to be calculated for the first time. The Smith Cloud's importance is that it provides insight into gas at a more advanced stage of accretion than the Magellanic Stream.

Earlier studies have shown how easily gas disrupts as it passes through a hot corona \citep{Moore94,Quilis01,Bland-Hawthorn07,Heitsch09}.
It is therefore remarkable that a massive, high-velocity gaseous system has survived so close to the disk.
In light of recent developments, we consider the prospect of a confining dark halo surrounding at least some HVCs to be plausible, as first proposed by \citet{Blitz01}, and therefore worthy of further investigation.
A number of ultra-faint dwarf galaxies have now been discovered in the Galactic halo \citep{Simon07,Kirby08,Geha09}.
What is particularly striking is that the inferred dark halo mass exceeds $\sim 10^7$~M$_\odot$ even for systems with very few stars \citep{Strigari08}.
These observations hint at the prospect of dark matter dominated galaxies containing \HI\ that have not experienced star formation. 
Complex H is one such `dark galaxy' candidate, with a mass dominated by dark matter if it is self-gravitating \citep{Simon06}.

\begin{figure}
  \center
  \includegraphics[width=0.45\textwidth]{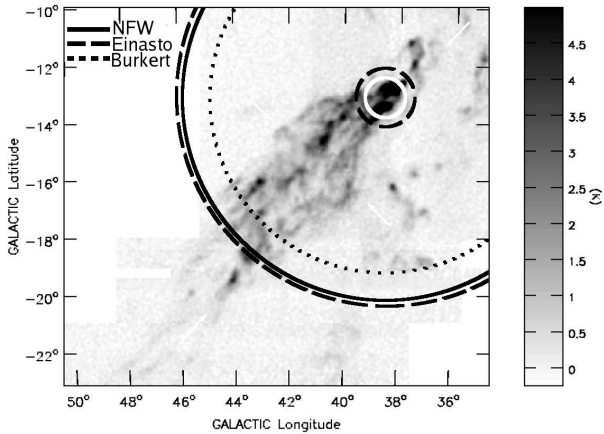}
  \caption{H{\scriptsize I} image of the Smith Cloud (Fig. 1. in \citet{Lockman08}) with superimposed core and 
  halo for three different dark matter models (NFW, Einasto, Burkert) discussed further in \S\ref{sec:MR}. 
  The tidally stripped mass today is $1-2\times10^8$~M$_\sol$ in dark matter. The core size assumes an initial gas 
  mass at apogalacticon of $1.1\times10^7$~M$_\sol$, although most of this gas is distributed along the orbit
  of the Smith Cloud today. We note that the Burkert profile has no surviving core, and the NFW core is displayed in white for visibility.}
  \label{fig:SC}
\end{figure}

Here we suggest that the Smith Cloud is another candidate for a dark galaxy, illustrated in Fig. \ref{fig:SC}, with the gas partly stabilised by a dark matter 
component.
We calculate the subhalo required for the survival of a gas cloud with properties similar to those observed in the Smith Cloud today.
In \S\ref{sec:DM} we discuss the properties of three possible dark matter profiles that may encompass the Smith Cloud.
In \S\ref{sec:MS} we discuss the orbit of the Smith Cloud within a realistic Galactic potential.
In \S\ref{sec:SAC} we present the equations that are used to calculate the ram pressure stripping of the subhalos along their orbit.
In \S\ref{sec:MR} we present the results of the semi-analytic models and discuss the implications of these results for the Smith Cloud.
In \S\ref{sec:conc} we summarize these results and discuss the limitations of our model.

\section{Dark Matter Profiles}\label{sec:DM}

Numerical simulations of dark matter halos indicate that they tend to be characterized by two parameters, typically a radial scale length and a density.
One of the first of these dark matter profiles was inspired by rotation curves in dwarf spiral galaxies \citep{Burkert95}, the `Burkert' profile, which incorporates a core of uniform density followed by a steady decline with increasing radius.
Early dissipationless N-body simulations however revealed that the dark matter core is more cusp-like.
The most widely used of the cuspy profiles is the `NFW' profile introduced by \citet{Navarro97}. But more recently, it seems that neither of these profiles are sufficiently flexible to describe what is now being seen in simulations involving 10$^{10}$ particles, with a more appropriate model provided by the Einasto profile \citep{Springel08}.

All of these profiles may be described by a density function $f_\rho(x)$ with the density at any point given by $\rho(x)=\rho_{ds}f_\rho(x)$, where $\rho_{ds}$ is the characteristic density.
Similarly the mass may be described by  $M(x) = M_{ds} f_m(x)$ and the potential by $\varphi(x,v_s) = v_s^2f_\varphi(x)$, where $v_s$ is the characteristic velocity of the halo.
The gas density for a hydrostatic cloud is given by $f_{gas}(x,v_s,c_g) \equiv n_H/n_{H,0}=\exp[-(v_s^2/c_g^2)f_\varphi(x)]$, where $c_g$ is the isothermal sound speed of the gas, and $n_{H,0}$ is the nuclear gas density.

These profiles are defined in terms of a scaled radius $x\equiv r/r_s$, where $r_s$ is a scale radius.
These parameters are defined by \citep{Sternberg02}:
\begin{eqnarray}
  M_{ds} &=& \frac{4}{3}\pi\rho_{ds}r_s^3, \\
  \rho_{ds} &=& \Delta\rho_u\frac{x^3_{vir}}{f_M(x_{vir})}, \label{eq:rhods}\\
  v_s &=& \left(\frac{4\pi}{3}G^3\Delta\rho_u\right)^{1/6}M_{vir}^{1/3}\left[\frac{x_{vir}}{f_M(x_{vir})}\right]^{1/2}, \label{eq:vs}\\
  r_s &=& \left(\frac{3}{4\pi\Delta\rho_u}\right)^{1/3}M_{vir}^{1/3}\frac{1}{x_{vir}},
\end{eqnarray}
where $\Delta$ is the overdensity of the dark matter $\sim340$ at $z=0$, $\rho_u = 2.8\times10^{-30}$~g~cm$^{-3}$ is the average density of the universe, $M_{vir}$ is the virial mass, and $x_{vir}$ is the scaled virial radius.
$x_{vir}$ is correlated, with a standard deviation $\sigma$, to the virial mass in the virial mass range of $10^8-10^{11}$~M$_\sol$ by \citep{Sternberg02}:
\begin{equation}
  x_{vir} = 27\times10^{0.14\sigma}\left(\frac{M_{vir}}{10^9 M_\sol}\right)^{-0.08} \label{eq:xvir}.
\end{equation}

\begin{figure}
  \center
  \includegraphics[angle=-90,width=0.49\textwidth]{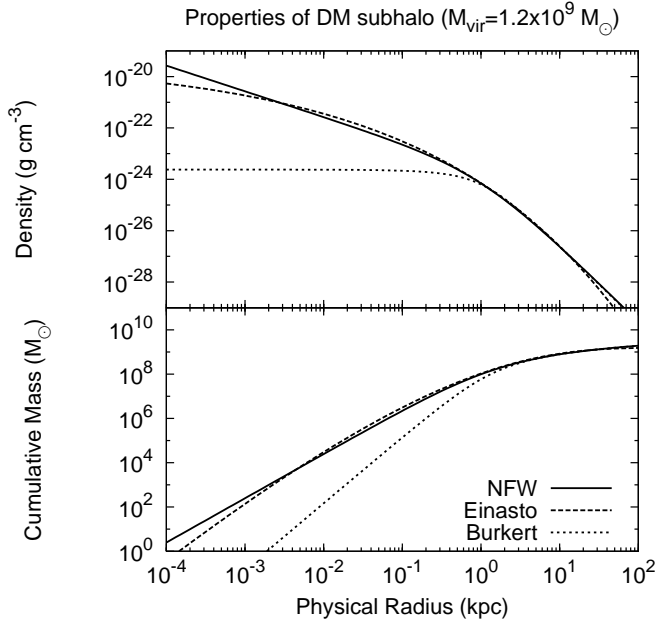}
  \caption{Comparison of the different subhalo density and mass profiles considered by this work prior to tidal stripping ($r_s=1.04$~kpc).}
  \label{fig:fm}
\end{figure}

Convenient forms for the NFW and Burkert profiles are given in \citet{Sternberg02}.
In Table 5 of that paper, we note that there is a typographical error in the NFW $f_{gas}$ equation which should read $f_{gas} = e^{-3(v_s/c_g)^2}(1+x)^{3(v_s/c_g)^2/x}$. 
In Table \ref{table:prof}, we give our formalism for the Einasto profile, and include the Burkert and (corrected) NFW profiles for completeness.
The three density and mass profiles are shown in Fig.~\ref{fig:fm} for median subhalos of virial mass $1.2\times10^{9}$~M$_\sol$, with corresponding scale radius $r_s=1.04$~kpc and profile dependent characteristics density determined by (\ref{eq:rhods}).

\begin{table*}
  \begin{centering}
  \begin{tabular}{|l|l|l|}
    \multicolumn{3}{c}{}\\
    \hline
    \hline
    NFW & Burkert & Einasto\\
    \hline
    $f_\rho(x) = x^{-1}(1+x)^{-2}$ & $f_\rho(x) = (1+x)^{-1}(1+x^2)^{-1}$ & $f_\rho = \exp[-2/\alpha(x^\alpha-1)]/4$\\
    $f_m(x) = 3\left[\ln(1+x)-\frac{x}{1+x}\right]$ & $ f_m(x) = \frac{3}{2}\left[\frac{\ln(1+x^2)}{2}+\ln(1+x)-\tan^{-1}x\right]$&$f_m = \beta\gamma(3/\alpha,2x^\alpha/\alpha)$\\
    $f_\varphi(x,v_s) = 3\left[1-\frac{\ln(1+x)}{x}\right]$ & $f_\varphi(x,v_s) =\frac{3}{2}[\left(1+\frac{1}{x}\right)\tan^{-1}x - \left(1+\frac{1}{x}\right)\ln(1+x)$&$f_\varphi(x,v_s) = \beta[2^{1/\alpha}\alpha^{-1/\alpha}\gamma(2/\alpha,2x^\alpha/\alpha)$\\
	&$\qquad+ \frac{1}{2}\left(1-\frac{1}{x}\right)\ln(1+x^2)]$ & $ \qquad- \gamma(3/\alpha,2x^\alpha/\alpha)/x-1]$\\
    $f_{gas}(x,v_s,c_g) = e^{-3(v_s/c_g)^2}(1+x)^{3(v_s/c_g)^2/x}$ & $f_{gas}(x,v_s,c_g) = [e^{-(1+1/x)\tan^{-1}x}(1+x)^{(1+1/x)}$ & $f_{gas}(x,v_s,c_g) = \exp(-v_s^2/c_g^2f_\varphi(x))$\\
      &$\qquad(1+x^2)^{(1/2)(1/x-1)}]^{(3/2)(v_s/c_g)^2}$&\\
    &&$\beta = (3/4)8^{-1/\alpha}e^{2/\alpha}\alpha^{-1+3/\alpha}$\\
    \hline
  \end{tabular}
  \caption{NFW, Burkert and Einasto profiles, where $\gamma$ is the lower incomplete gamma function, $x\equiv r/r_s$ is the scale radius, $v_s$ is the halo circular velocity and $c_g$ is the gas sound speed; cf. \citet{Sternberg02}, Table 5\label{table:prof}. The four quantities in each column are the density profile $f_p$, the dark matter mass profile $f_m$, the dark matter potential profile $f_\varphi$, and the gas density profile $f_{gas}$.}
  \end{centering}

\end{table*}

For dark matter subhalos moving within the Galactic environment, the radial profile will be altered in two competing ways.
The condensation of baryons contracts the inner core of the subhalo while ram-pressure stripping leads to an expansion of the subhalo.
The overall impact of adiabatic contraction can be calculated iteratively using \citep{Blumenthal86}:
\begin{equation}
  r\left[M_b(r)+M_{DM}(r)\right] = r_iM_i(r_i),
\end{equation}
where $M_b(r)$ is the final baryon mass distribution, $M_{DM}(r)$ is the final dark matter distribution and $M_i(r_i)$ is the initial total mass distribution at an initial orbital radius $r_i$.
By assuming the orbits of baryonic material do not cross, then $M_{DM}(r) = (1-F)M_i(r_i)$ where $F$ is the baryon to dark matter
fraction. We have investigated the impact of adiabatic contraction and find that it has a negligible influence in modelling the Smith Cloud.

We note that the virialization time of the dark matter halo can also be important.
Here the overdensity of the dark matter compared to the average density is given by \citep{Bryan98}:
\begin{equation}
  \Delta = \frac{18\pi^2 + 82x-39x^2}{\Omega(z)},
\end{equation}
where $x = \Omega(z)-1$, and for a flat universe
\begin{equation}
  \Omega(z) = \frac{\Omega_m(1+z)^3}{\Omega_m(1+z)^3+\Omega_\Lambda}.
\end{equation}
The physical density of a halo is directly proportional to this overdensity factor such that $\rho_{ds} = \rho_o\Delta/\Delta_o$ where
$\rho_o$ is the central density of a halo that virialized at $z=0$. This factor also contributes to other halo properties such as the scale radius $r_s \propto \Delta^{-1/3}$ and the scale velocity $v_s \propto \Delta^{1/6}$.

\section{Model Setup}\label{sec:MS}

We consider two models of evolution, one in which the Smith Cloud is infalling for the first time, hereafter the Infalling Orbit Models,
and a second model where the Smith Cloud has already been maximally stripped due to previous orbits, hereafter the Repeated
Orbit Models. These both share common features: (i) they have the same trajectory today, (ii) the dark matter halo has been tidally 
stripped down from some larger initial mass ($M_{\mathrm{vir}}$) in an identical fashion before our calculations commence at apogalacticon. 
The important distinction is tidal stripping of
the gas is possible in the Infalling Orbit Models but not in the Repeated Orbit Models; in both cases, ram pressure stripping by the
hot halo is important. For each case, the evolution of the Smith Cloud is considered for the NFW, Einasto and Burkert models.

The evolution of the model clouds was calculated as a function of three variables: the initial virial mass at the time of formation (i.e. before
the dark-matter halo fell into the Galaxy), the dark matter profile at this time, and the initial hydrogen gas mass at apogalacticon. 
For both the Repeated Orbit Model and Infalling Orbit Models, the evolution of $7503$ model clouds were calculated, corresponding to $61$ logarithmically spaced virial masses in the range $M_{\mathrm{vir}}=5\times10^7-5\times10^{10}$~M$_\sol$ and $41$ logarithmically spaced 
gas masses in the range $M_{\mathrm{gas}}=1\times10^6-1\times10^8$~M$_\sol$.

\begin{figure}
  \center
  \includegraphics[width=0.5\textwidth, bb= -100   220   760   580, clip=true]{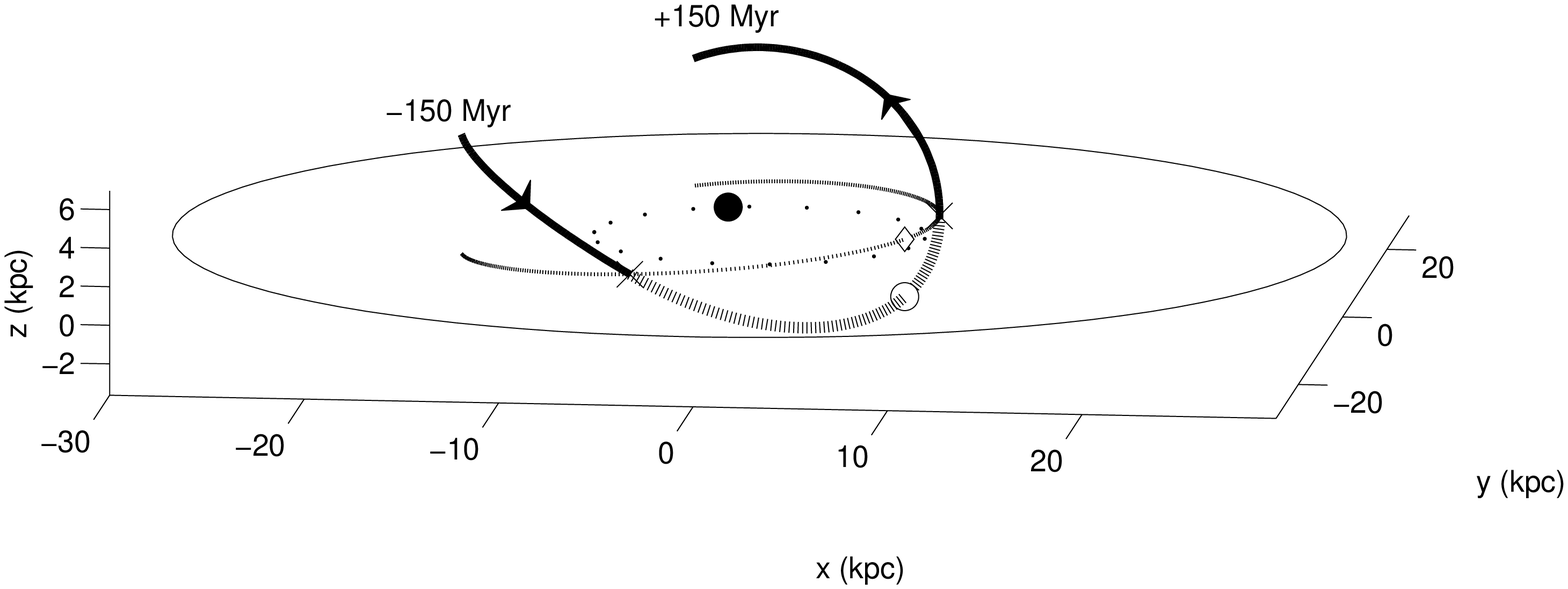} 
  \caption{The orbit of the Smith Cloud, calculated using the potential from \citet{Wolfire95}. The current position is represented by an unfilled circle and the Smith Cloud is travelling in the direction of the arrows, with heights below the disk represented by a dotted line. The Sun's position is shown as a filled circle on the Solar Circle. The thin dotted line represents the projection of the Smith Cloud onto the disk. The disk is represented by a solid line at $30$~kpc.}
  \label{fig:orb}
\end{figure}

The orbit of the Smith Cloud was calculated using the position and velocity data from \citet{Lockman08} for the tip of the Smith Cloud: 
$(R,z)=(7.6,-2.9)$~kpc, $(v_{R},v_{\phi},v_z) = (94,270,73)$~km~s$^{-1}$. The form of the Galactic potential is given by \citet{Wolfire95} 
normalized  by a  circular velocity of $v_c = 220$~km~s$^{-1}$ at the Solar Circle. In Fig. \ref{fig:orb}, we show the predicted orbit of the cloud system. In agreement with Lockman et al (2008), we find that the Smith Cloud has intersected the disk $\sim70$~Myr ago and will pass through the disk again in $\sim30$~Myr. 

For all subhalo models, we investigate the effects of dynamical friction on the orbit trajectory.
The formalism used is described by \citet{Jiang00}: we point out that the value for the circular velocity in their Table 1 should be $v_c = 235$ km s$^{-1}$ (not 181 km s$^{-1}$ as quoted) to be consistent with their analysis.
But over the past few hundred million years, dynamical friction is found to have only minimal effect, even in the high mass limit.
This is because, once again, the impact of gas loss from the subhalo close to the disk is found to dominate the evolution of the subhalo.
We assume that any drag between the model clouds and the Galactic corona is negligible and does not affect the orbit.

Each model cloud is considered to be a dark matter potential well filled with gas in isothermal hydrostatic equilibrium.
We assume a primordial helium abundance $n_{He}/n_{H} = 1/12$ and metallicity of $Z/Z_\sol=0.1$.
We also assume the gas has a temperature of $1.2\times10^{4}$~K and adopt an ionization fraction of $50\%$ for the Smith Cloud, slightly below the newly updated H$^+$/H$^0$ ratio in \citet{Hill09}.
This temperature and ionization fraction then give a sound speed of $c_g=11$~km~s$^{-1}$.
The gas is distributed in the potential well according to the gas density profile $n_H(x,v_s,c_g)=n_{H,0}f_{gas}(x,v_s,c_g)$, where $x\equiv r/r_s$ is the scale radius, $v_s$ is the halo circular velocity given by (\ref{eq:vs}) and $f_{gas}(x,v_s,c_g)$ is given in Table \ref{table:prof}.

For the initial dimension of the model clouds, the sound crossing time is $200$~Myr, falling to about $30$~Myr at the disk.
We therefore begin each orbit at the apogalacticon, approximately $17$~kpc from the Galactic centre, which took place $150$~Myr ago, and $\approx 100$~Myr after the last transit of the disk.
We hence assume the cloud has regained hydrostatic equilibrium by this point and is the basis for our initial assumption of hydrostatic equilibrium.
Due to the comparatively large sound crossing time we assume that the radial gas density profile does not change throughout the calculation, only being truncated, however, due to the lower sound crossing time at the disk we use a more restrictive condition as the cloud transits the disk (\S\ref{sec:SAC}).

In order to trace the long-term impact of ram-pressure stripping, we must define the density profile of the hot coronal halo and the
Galactic gas disk. The acceleration due to the stellar density profile is already included in the Wolfire potential.
The Galactic corona is assumed to be a mix of H$+$He at $T_h=2\times10^6$~K with the density calculated from the gravitational potential of \citet{Wolfire95}, normalized to a plasma density of $n_{H,0}=10^{-4}$~cm$^{-3}$ at $55$~kpc from the Galactic Centre \citep{Bland-Hawthorn08}. The density profile of the \HI\ disk was calculated using an exponential drop-off in density both in the radial and
vertical direction. We use the new prescription in \citet{Kalberla08} which includes the vertical flaring of the outer disk.

\section{Analytic Procedure for Cloud/Halo Evolution}\label{sec:SAC}

The Smith Cloud was evolved through a procedure that iteratively calculates the competing processes on the dark matter 
confined cloud. The initial radius of the gas comes from integrating the gas density profile over space giving 
\citep[modified from][]{Sternberg02}:
\begin{equation}
  M_{gas}(0) = 4\pi{}r_s^3m_Hf_{gas}(x_0)\frac{P(0)}{1.583k_bT_c}\int^{x_0}_0 x^2f_{gas}(x) dx, \label{eq:Mass}
\end{equation}
where $r_s$ is the scale radius of the halo, $x_0=r(0)/r_s$, $f_{gas}(x)=f_{gas}(x,v_s,c_g)$ is the gas distribution profile for constant scale velocity and sound speed.
Here we have replaced $n_{H,0}$ with $P/(nk_bT)f_{gas}(x_0)$ by assuming that initially the outer edge of the halo is bound by thermal pressure from the Galactic halo.

At each step we calculate whether the gas at the cloud edge is prone to stripping by examining if the force applied by the external medium exceeds the maximum gravitational force in the subhalo at a given radius. For an isothermal spherical cloud, gas will be stripped if \citep{McCarthy08}:
\begin{equation}
  \rho_{gal}v^2_{c} > \frac{\pi{}GM(R)\rho_{c}}{2R},
\end{equation}with $\rho_{gal}$ and $\rho_{c}$ being the density of the Galactic gas and the outer subhalo gas respectively, $v_{c}$ the relative velocity between these two gases, and $M(R)$ the amount of mass within a radius $R$.
The mass within a radius $R$ comprises the dark matter subhalo $M_{DM}$ and the gas within that radius $M_{gas}$.

If this stripping criterion is satisfied, then a shock passes through the gas, at a speed given by \citep{McCarthy08}:
\begin{equation}
  v_{fs} = \frac{4}{3}v_{c}\sqrt{\frac{\rho_{gal}}{\rho_{c}}}.
\end{equation}
This gas assumed to be no longer bound and is removed from the cloud.
By assuming that the shock removes a shell of gas, the outer radius of the gas changes at a rate given by
$\dot{r} = -v_{fs}/2$.
As $v_{fs}$ tends to be much larger than the gas sound speed, $c_g$, we assume that the density profile of the gas 
does not change, becoming truncated at the edge of the shock.

For Infalling Models, the tidal stripping must also be considered.
At each step, the tidal radius and tidal mass of the subhalo are also calculated. The tidal radius is given by \citep{Hayashi03}:
\begin{equation}
  \frac{M(r_t)}{r_t^3} = \left[2-\frac{R}{M_{gal}(R)}\frac{\partial{}M_{gal}}{\partial{}R}\right]\frac{M_{gal}(R)}{R^3},
\end{equation}
where $R$ is the distance between the centre of the subhalo and the Galactic centre.
Any gas outside of the tidal radius $r_t$ will also be removed, with the new gas radius being the minimum of the tidal radius or the radius resulting from ram pressure stripping.

The total mass can then be calculated by modifying (\ref{eq:Mass}) such that
\begin{equation}
  M_{gas}(t) = 4\pi{}r_s^3m_Hf_{gas}(x_0)\frac{P(0)}{1.583k_bT_c}\int^{x(t)}_0 x^2f_{gas}(x) dx,
\end{equation}
where $x(t) = r(t)/r_s$. The central column density is calculated using
\begin{equation}
  N_{H,c}(t) = 2\frac{P(0)}{1.583k_bT_c}\int^{x(t)}_0 f_{gas}(x) dx.
\end{equation}
We iterate through the orbit in $1$~Myr steps to determine the evolution of radius, mass and column density for the cloud/halo system.
Convergence was tested by iterating through on $0.5$~Myr steps for some surviving clouds, with the smaller step size producing less than $1\%$ difference in mass calculations throughout the orbit.

A special step, however, is considered when the Smith Cloud is 
moving through the Galactic disk.
Consider a cloud with a gas density $\rho_c$, radius $r_c$ moving at 
speed $v_c$ through the Galactic halo. Consider a gaseous disk where, 
at some radius, it has a local gas density
$\rho_d$ and a half-scale height $h_d$. Will the cloud survive
its transit through the disk?

We write down two shock conditions: one for the shock
driven into the cloud by the disk, and one for the shock driven
by the cloud into the disk. The cloud survives if the shock through
the cloud takes longer to propagate than the shock through the disk.
The shock propagates into the disk gas at speed 
\begin{equation}
v_{s,d} = {4 v_c \over 3 (1 + \sqrt{\rho_d / \rho_c})}
\end{equation}
and another into the cloud, at speed 
\begin{equation}
v_{s,c} = {4 v_c \sqrt{\rho_d / \rho_c}\over 3 (1 + \sqrt{\rho_d / \rho_c})}
\end{equation}
The shock needs to be fast enough to heat the disk gas significantly.

The key to punching a hole is that the shock must go right through the
disk gas before passing through the cloud.  In other words we need
roughly $r_c/v_{s,c} > h_d/v_{s,d}$ or equivalently $v_{s,d}/v_{s,c} > h_d/r_c$.
This leads to
\begin{equation}
\sqrt{\rho_c / \rho_d} > h_d/r_c
\end{equation}
It follows that, to first order, the cloud will survive if
\begin{equation}
  N_{H,c} > \frac{h_{d}}{r_c}N_{H,d},
\end{equation}
where $N_{H,c}$ and $N_{H,d}$ are the column densities of the cloud and the disk along the orbit respectively. This is the kinetic energy argument presented by \citet{Bland-Hawthorn08}.
We note that the dark matter
subhalo itself has negligible impact on the Galactic gas.

We therefore expect that all gas in the Smith Cloud at a column density less than $h_{d}/r_c\cdot N_{H,d}$ is stripped when the cloud moves through the disk.
As the passage through the disk occurs on a short timescale $\la1$~Myr, we treat this removal of gas as instantaneous, and in addition to the normal ram pressure stripping that occurs during that time step. 
We also set our model disk to rotate with equal angular velocity to the Smith Cloud as it passes through at $\approx 13$~kpc, and a static disk otherwise.
Although this choice is unphysical and in both cases contradict, the observed relative velocity between the Smith Cloud and the ISM, the cloud spends only a short time near the disk, and hence any additional gas stripped via ram pressure will contribute only a small correction to the mass lost from the cloud near the disk.

\section{Model Results}\label{sec:MR}
\begin{figure*}
  \center
  \includegraphics[width=1\textwidth]{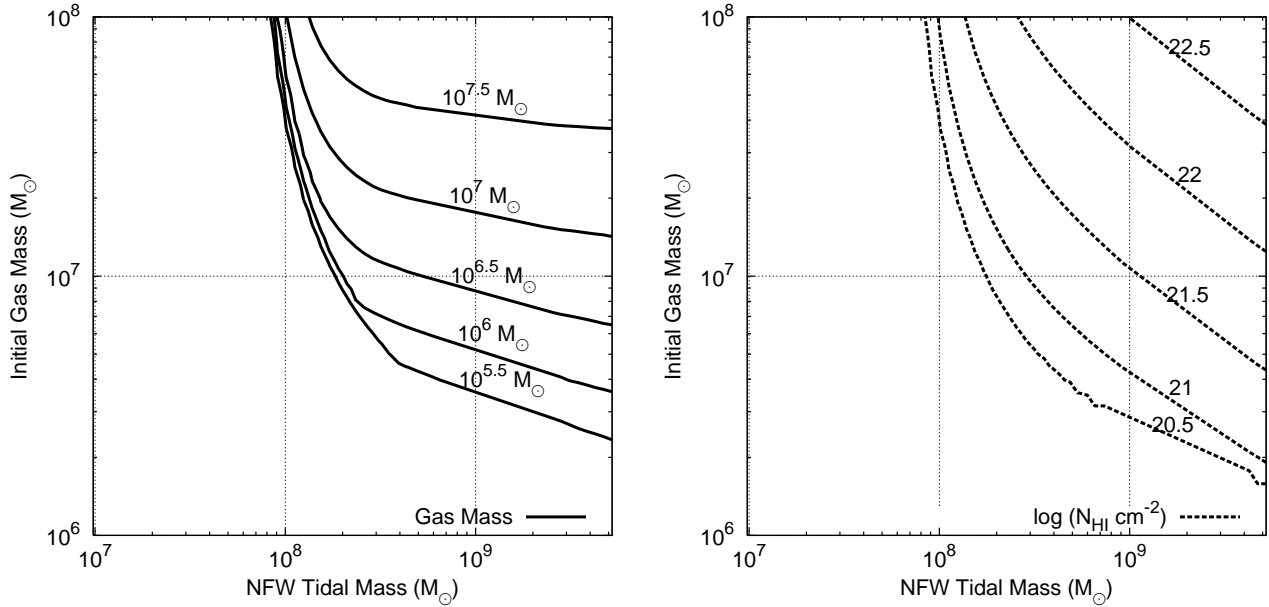}
  \caption{Final gas mass and column density contours for the Repeated Orbit NFW model. Here the gas mass refers only to hydrogen to facilitate comparison with observations; an extra factor of 1.33 is needed to determine the total gas mass.}
  \label{fig:NFWcont}
\end{figure*}

\begin{figure*}
  \center
  \includegraphics[width=\textwidth,angle=-90, bb=-250 100 554 650]{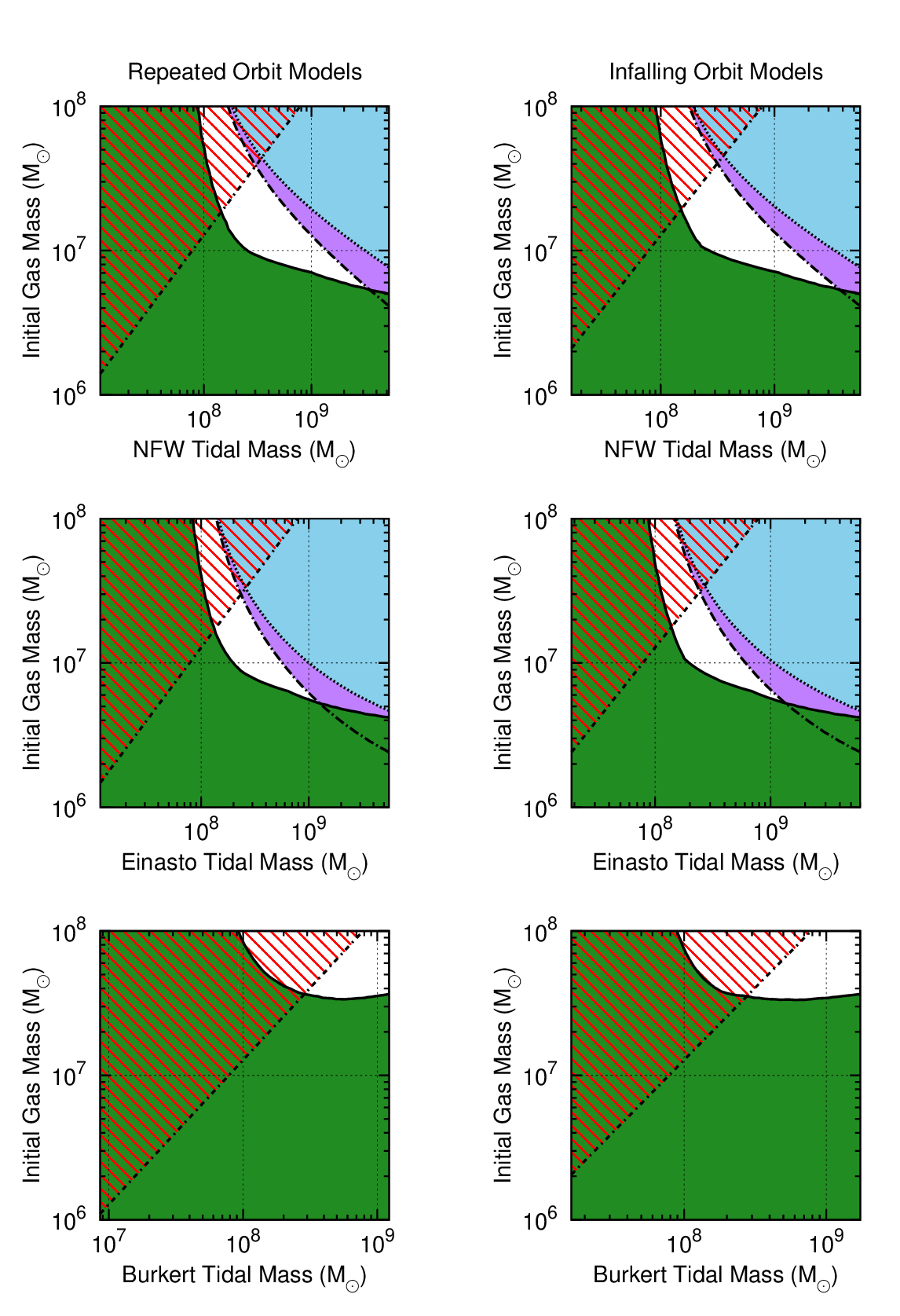}
  \caption{The initial gas mass vs. the surviving tidal mass for three different dark matter models. The left hand column shows models which will have already been reduced to the final tidal radius by the beginning of the simulation; the right hand column shows models where tidal stripping is occurring at the same time as ram pressure stripping, simulating a cloud infalling for the first time. The allowed solutions for the Smith Cloud are shown in white. The upper and lower limits are discussed in \S \ref{sec:MR}.} 
\label{fig:Smith}
\end{figure*}

In our simulations, a large number of model clouds survive to the present day, although not all of these are 
similar to the Smith Cloud as observed today. We interpret our results graphically through contour plots of properties derived from the mass, radius and column density of the model clouds. The simplest properties
$-$ mass, column density $-$ are displayed for the Repeated Orbit Model NFW profile in Fig. \ref{fig:NFWcont}.
By removing model results which are greatly dissimilar to those observed for the Smith Cloud, we can find an allowed parameter space for the initial cloud properties (Fig. \ref{fig:Smith}). In Figs. 4-7, we present our results in terms of the mass of the tidally-stripped subhalo rather than the
mass of the initial infalling halo.

The lower limit of the allowed parameter space is found by requiring the amount of gas in the model clouds to match the observations.
The Smith Cloud has an \HI\ mass of at least $10^6$~M$_\sol$ \citep{Lockman08}, and at least as much again in H{\small II} \citep{Hill09}, giving a final gas mass of at least $2\times10^6$~M$_\sol$.
The excluded region corresponding to lower cloud gas masses is shown in green (Fig. \ref{fig:Smith}) bounded by a solid black line.

We include several upper bounds to the allowed parameter space.
The universal baryonic to dark matter mass fraction is $\Omega_b/\Omega_m\approx 0.17$ from the latest microwave background 
observations.\footnote{Throughout the paper, we assume a $\Lambda$CDM cosmology with parameters ($\sigma_8, h, \Omega_b, \Omega_m, \Omega_\Lambda) = (0.8, 0.7, 0.04, 0.3, 0.7).$}
While there may be galaxies that exceed this limit (e.g. dwarf galaxies that form in tidal tails of strong mergers), it is reasonable to assume that on average most dark matter subhalos do not accrete more than this fraction.
Thus, we can probably exclude all clouds with baryon to dark matter mass fractions above this limit, displayed by the red shaded region bounded by a dot-dash line in Fig. \ref{fig:Smith}.

In dwarf galaxies, the highest \HI\ column densities seen outside of star-forming regions are always less than $2\times10^{21}$~cm$^{-2}$. This critical column density is found in regions with $\le{}1\sigma$ of optical emission (B. Ekta 2009, private communication).
Since no stars are observed in the Smith Cloud, we include this upper limit to the \HI\ column density, shown by the light blue region bounded by a dotted line in Fig. \ref{fig:Smith}.

An extensive literature demonstrates that star formation is intimately associated with the presence of molecular gas \citep[e.g.][]{Calzetti09}.
For a canonical HVC metallicity of about $0.1$~Z$_\sol$, it is unlikely that molecular gas can be measured directly from the Smith Cloud
with existing submillimetre telescopes \citep{Ritcher01}. This sets a stronger, but presumably less rigorous constraint on the allowed regions 
of the Smith Cloud; the excluded regions are shown in purple bounded by a dot-dash line in Fig. \ref{fig:Smith}.
The detection of molecular gas in the Smith Cloud would alter this limit from an upper limit, to a lower limit and hence alter the allowed parameter space to exist only in the purple region.
All of the upper limit bounds shown in Fig. \ref{fig:Smith} have a strong dependence on the gas distribution within the model.
Here we have assumed that the gas was initially in hydrostatic equilibrium with the profile only being truncated and not altered further; 
shocks that substantially modify the profile are associated with gas that gets stripped.

The NFW and Einasto profiles restrict the allowed regions of the cloud to a narrow range of initial gas masses and subhalo masses, with the subhalo mass range restricted to $10^8-10^9$~M$_\sol$ and gas mass initially $5\times10^6-5\times10^7$~M$_\sol$, although we note that the allowed gas mass range is much larger at smaller subhalo masses than at the higher mass end.
By comparison to the gas mass present today, the Smith Cloud will have lost between $50\%$ and $95\%$ of the gas present at apogalacticon.
A large fraction of this gas loss in all model clouds is due to the transit of the disk suggesting, suggesting that the Smith Cloud may not survive the next passage through the disk.
The tidal stripping of dark matter does not substantially affect the gas contained in the model clouds, with the gas contained within the protected core of the potential well.
In the Infalling Orbit models, the Smith Cloud will also have lost approximately $50\%$ of the dark matter mass as well to tidal stripping, although most of this mass loss will have occurred before the semi-analytic calculations were begun.
This scenario of an HVC infalling for the first time, possibly after interacting with a third body, such as a dwarf galaxy, has the potential to increase the baryonic to dark matter ratio above the universal fraction, particularly if the orbit avoided dense sections of the disk.
The Burkert profile retains a lower column of gas, due to a lower central density, with the effect that the region is not as well constrained as in the case of a NFW or a Einasto profile.
It is this lower column that leads to a higher mass requirement in order to survive traversal of the disk, which is dictated by the column density of the cloud.

\begin{figure}
  \center
  \includegraphics[angle=-90,width=0.5\textwidth]{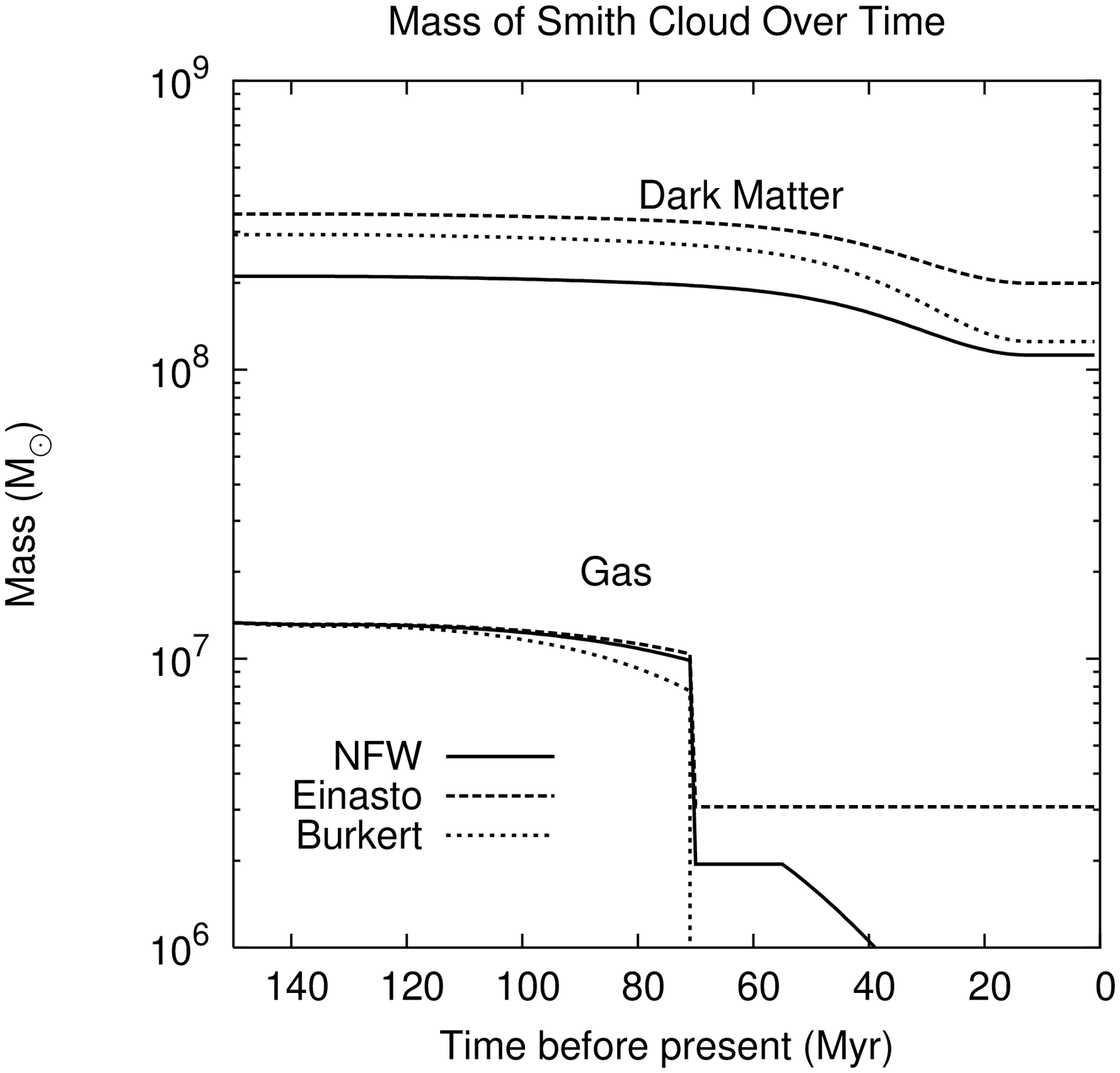}
  \caption{The change in tidal dark matter and baryonic mass for a model over the last $150$~Myr with input parameters (M$_{\mathrm{vir}}=1.2\times10^{9}$~M$_\sol$ [M$_{\mathrm{tidal}}\approx3\times10^8$~M$_\sol$], M$_{\mathrm{gas}}$=$1.1\times10^7$~M$_\sol$). The dark matter subhalos begin the simulation with a tidal mass of approximately $3\times10^8$~M$_\sol$ and a hydrogen mass of $1.1\times10^7$~M$_\sol$.}
  \label{fig:massloss}
\end{figure}

We display the mass history of three Infalling Orbit Model clouds corresponding to the same input parameters for each dark-matter profile 
(M$_{\mathrm{vir}}=1.2\times10^{9}$~M$_\sol$ [M$_{\mathrm{tidal}}\approx3\times10^8$~M$_\sol$], M$_{\mathrm{gas}}$=$1.1\times10^7$~M$_\sol$) in Fig. \ref{fig:massloss}.
The hydrostatic pressure bound gas radius is located within the tidally protected core of the dark matter subhalo, and therefore the mass history of the Repeated Orbit Model with the same initial conditions will be similar. 

\begin{figure}
  \center
  \includegraphics[angle=-90,width=0.5\textwidth]{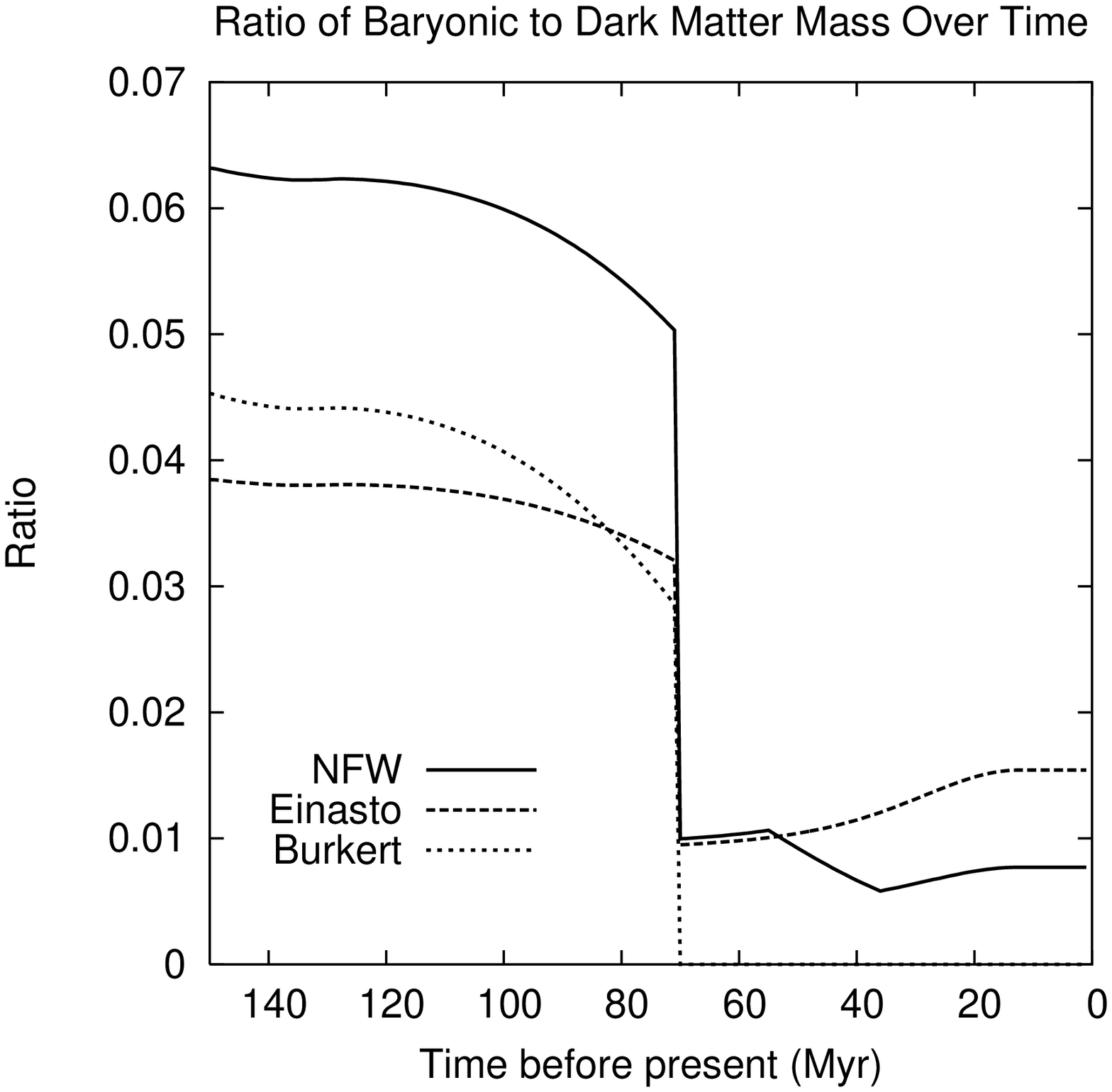}
  \caption{The change in baryonic to tidal dark matter mass ratio for a model over the last $150$~Myr with input parameters (M$_{\mathrm{vir}}=1.2\times10^{9}$~M$_\sol$ [M$_{\mathrm{tidal}}\approx3\times10^8$~M$_\sol$], M$_{\mathrm{gas}}$=$1.1\times10^7$~M$_\sol$). The dark matter subhalos begin the simulation with a tidal mass of approximately $3\times10^8$~M$_\sol$ and a hydrogen mass of $1.1\times10^7$~M$_\sol$. \label{fig:BDMratio}}
\end{figure}

As the subhalo loses dark matter to tidal stripping and gas to ram pressure stripping at different rates, the baryon to dark matter fraction will alter over the course of the orbit.
This change, displayed in Fig. \ref{fig:BDMratio} for the Infalling Orbit Model clouds corresponding to the input parameters (M$_{\mathrm{vir}}=1.2\times10^{9}$~M$_\sol$ [M$_{\mathrm{tidal}}\approx3\times10^8$~M$_\sol$], M$_{\mathrm{gas}}$=$1.1\times10^7$~M$_\sol$) is particularly apparent at the transit of the disk where a large amount of gas is lost but the dark matter component remains unchanged.
Noting that the dark matter halo will have already been tidally stripped at the beginning of the evolutionary model suggests that HVCs may undergo radical changes in their baryon to dark matter fraction along their orbits.

\section{Summary}\label{sec:conc}
We have modelled the Smith Cloud by assuming that it is supported against total disruption by a dark-matter subhalo.
By considering the evolution of this system over the past $150$~Myr, 
we are able to compare our predicted properties to the known properties of the Smith Cloud today.
Under the assumption of dark-matter confinement, the subhalo encompassing the Smith Cloud has a tidal mass of 
$\sim 3\times10^8$~M$_\sol$ at the present time. Even before this subhalo arrived at its present orbit, it will have
lost $50-90$\% of its dark matter and a comparable amount of progenitor gas due to tidal and ram pressure stripping by the Galaxy.
In this respect, the likelihood of an infalling (dark-matter confined) ``Smith Cloud'' event is comparable to the rate of dwarf galaxy infall, i.e. roughly one
event per Gyr.

If the NFW and Einasto profiles are appropriate descriptions of dark matter halos, we find that the Smith Cloud can
only exist in a narrow strip of the available parameter space defined by the initial dark matter subhalo mass and
the initial gas mass. This remains true with and without dark matter stripping.
This narrow strip also appears, albeit shifted with respect to the present axes, if the Galactic halo is considered to be 
$10\times$ over or underdense from our model setup, or if the dark matter containing subhalo was virialized at an 
earlier redshift and correspondingly is denser than a subhalo virialized today.

Consistent with our picture, the observed wake in Fig. 1 presumably resulted from
ablation processes due to the impulsive shock of the disk transit. We note with interest that the wake flares
(i.e. expands transverse to the direction of the wake) beyond the putative dark matter core (Fig. 1). 
In our picture, this can be understood in the context of a transsonic flow. The Smith Cloud is moving at up to 300\kms\ with respect
to a hot halo with adiabatic sound speed $\sim$200\kms\ making the cloud-halo interface mildly supersonic. Once the stripped gas
is left behind, it must undergo mixing with the hot halo through Kelvin-Helmholtz instabilities, causing the stream to expand perpendicular 
to the flow. The temperature of the \HI\ gas is predicted to rise along the wake. Unless the confining core-dominated, dark matter halo is 
very massive, we do not expect the declining potential seen by the wake material to influence the degree of flaring. These issues
are being addressed in new hydrodynamical simulations of the Smith Cloud that we will present elsewhere.

With respect to further constraints on shock interactions, we expect that
shock signatures at UV to x-ray wavelengths will have largely faded away, and the
\HI\ ``hole'' at the crossing point will have been substantially stretched by
differential shear. For cloudlets smaller than 100 pc, thermal 
conduction due to the halo corona \citep{McKee77} and the Galactic radiation 
field convert the ablated gas to a clumpy plasma. We therefore strongly encourage more extensive
mapping of the \HI\ and the ionized gas.

As we discuss elsewhere \citep{Bland-Hawthorn08}, it is very difficult to see how the Smith Cloud, like several other
large HVCs, could have come in from, say, the distance of the Magellanic Stream. This was our motivation
for considering a cloud stabilised against disruption by a confining dark matter halo.
In essence, we have considered the prospect that the Smith Cloud constitutes a `dark galaxy' where
star formation never took place. 
Any evidence of a co-moving stellar population would have profound consequences.
An essential requirement for star formation to occur appears to be
the presence of molecular gas \citep[e.g.][]{Calzetti09}. 
We envisage much deeper searches for the presence of molecular gas in compact HVCs, in particular, 
with the newly commissioned {\em Cosmic Origins Spectrograph} on the {\em Hubble Space Telescope}.
 
Our model is not unique.
The cloud may have been dislodged from the outer disk or confining potential of an infalling dwarf galaxy.
A cloud metallicity of [Fe/H]$\approx$-1 may be appropriate in either scenario but there are no obvious candidates at the present time.
The interloper must be on a prograde orbit which rules out some
infalling dwarfs \citep[e.g. $\omega$Cen;][]{Bekki03}
but conceivably implicates disrupting dwarfs like Canis Major, assuming these were still losing gas in the recent past.

\acknowledgments MN is supported by an Australian Postgraduate Award. JBH is supported by a Federation Fellowship
from the Australian Research Council. We are indebted to the referee for a very constructive and well considered report.
In addition we are grateful to Ralph Sutherland for his helpful comments on the paper.


\begin{thebibliography}{}
\bibitem[{Bekki \& Freeman}, 2003]{Bekki03}Bekki, K., \& Freeman, K.~C. 2003, \mnras, 346, L11
\bibitem[{Bland-Hawthorn et al.}, 1998]{Bland-Hawthorn98}Bland-Hawthorn, J., Veilleux, S., Cecil, G.~N., Putman, M.~E., Gibson, B.~K. \& Maloney, P.~R. 1998, \mnras, 299, 611
\bibitem[{Bland-Hawthorn et al.}, 2007]{Bland-Hawthorn07}Bland-Hawthorn, J., Sutherland, R., Agertz, O., \& Moore, B. 2007, \apj, 670, L109
\bibitem[{Bland-Hawthorn}, 2008]{Bland-Hawthorn08}Bland-Hawthorn, J. in IAU Symp. 254, The Galaxy Disk in Cosmological Context, ed. J. Andersen, J. Bland-Hawthorn, \& B. Nordstr{\"o}m (Cambridge: Cambridge Univ. Press), 241
\bibitem[{Blitz et al.}, 2001]{Blitz01}Blitz, L., Spergel, D.~N., Teuben, P.~J., Hartmann, D. \& Burton, W.~B. 1999, \apj, 514, 818 
\bibitem[{Blumenthal et al.}, 1986]{Blumenthal86}Blumenthal, G.~R., Faber, S.~M., Flores, R., \& Primack, J.~R. 1986, \apj, 301, 27
\bibitem[{Bryan \& Norman}, 1998]{Bryan98}Bryan, G., \& Norman, M. 1998, \apj, 495, 80
\bibitem[{Burkert}, 1995]{Burkert95}Burkert, A. 1995, \apj, 447, L25
\bibitem[{Calzetti \& Kennicutt}, 2009]{Calzetti09}Calzetti, D., \& Kennicutt, R.~C. 2009, arXiv:0907.0203
\bibitem[{Ekta, Chengalur, \& Pustilnick}, 2008]{Ekta08}Ekta, Chengalur, J.~N., \& Pustilnick, S.~A. 2008, \mnras, 391, 881
\bibitem[{Geha et al.}, 2009]{Geha09}Geha, M., Willman, B., Simon, J.~D., Strigari, L.~E., Kirby, E.~N., Law, D.~R., \& Strader, J. 2009, \apj, 692, 1464
\bibitem[{Hayashi et al.}, 2003]{Hayashi03}Hayashi, E., Navarro, J.~F., Taylor, J.~E., Stadel, J., \& Quinn, T. 2003 \apj, 584, 541
\bibitem[{Heitsch \& Putman}, 2009]{Heitsch09}Heitsch, F., \& Putman, M.~E. 2009, \apj, 698, 1485
\bibitem[{Hill, Haffner \& Reynolds}, 2009]{Hill09}Hill, A.~S., Haffner, L.~M., \& Reynolds, R.~J. 2009, \apj, 703, 1832
\bibitem[{Jiang \& Binney}, 2000]{Jiang00}Jiang, I., Binney, J. 2000, \mnras, 314, 468
\bibitem[{Kalberla \& Dedes}, 2008]{Kalberla08}Kalberla, P.~M.~W., \& Dedes, L. 2008, \aap, 487, 951
\bibitem[{Kirby et al.}, 2008]{Kirby08}Kirby, E.~N., Simon, J.~D., Geha, M., Guhathakurta, P., \& Frebel, A. 2008, \apjl, 685, L43 
\bibitem[{Lockman}, 1984]{Lockman84}Lockman, F.~J. 1984, \apj, 283, 90 
\bibitem[{Lockman et al.}, 2008]{Lockman08}Lockman, F.~J., Benjamin, R.~A., Heroux, A.~J., \& Langston, G.~I. 2008, \apjl, 679, L21
\bibitem[{Mazure, \&  Capelato}, 2002]{Mazure02}Mazure, A., \& Capelato, H.~V. 2002, \aap, 383, 384
\bibitem[{McCarthy et al.}, 2008]{McCarthy08}McCarthy, I.~G., Frenk, C.~S., Font, A.~S., Lacey, C.~G., Bower, R.~G., Mitchell, N.~L., Balogh, M.~L., \& Theuns, T. 2008, \mnras, 383, 593
\bibitem[{McKee \& Cowie}, 1977]{McKee77}McKee, C.~F., \& Cowie, L.~L. 1977, \apj, 215, 213
\bibitem[{Moore \& Davis}, 1994]{Moore94}Moore, B., \& Davis, M. 1994, \mnras, 27, 209
\bibitem[{Navarro, Frenk \& White}, 1997]{Navarro97}Navarro, J.~F., Frenk, C.~S., \& White, S.~D.~M 1997, \apj, 528, 607
\bibitem[{Putman et al.}, 2003]{Putman03}Putman, M.~E., Bland-Hawthorn, J., Veillux, S., Gibson, B.~K., Freeman, K.~C., \& Maloney, P.~R. 2003, \apj, 597, 948
\bibitem[{Quilis \& Moore}, 2001]{Quilis01}Quilis. V., \& Moore, B. 2001, \apjl, 555, L95
\bibitem[{Ritcher et al.}, 2001]{Ritcher01}Richter, P. and Sembach, K.~R. and Wakker, B.~P. and Savage, B.~D. 2001, \apjl, 562, L181
\bibitem[{Simon et al.}, 2006]{Simon06}Simon, J.~D., Blitz, L., Cole, A.~A. Weinberg, M.~D., Cohen, M. 2006, \apj, 640, 270
\bibitem[{Simon \& Geha}, 2007]{Simon07}Simon, J.~D., \& Geha, M. 2007, \apj, 670, 313
\bibitem[{Smith}, 1963]{Smith63}Smith, G.~P. 1963, Bull. Astron. Inst. Netherlands, 17, 203
\bibitem[{Springel et al.}, 2008]{Springel08}Springel, V., Wang, J., Vogelsberger, M., Ludlow, A., Jenkins, A., Helmi, A., Navarro, J.~F., Frenk, C.~S., \& White, S.~D.~M. 2008, \mnras, 391, 1685
\bibitem[{Sternberg, McKee \& Wolfire}, 2002]{Sternberg02}Sternberg, A., McKee, C.~F. \& Wolfire, M.~G. 2002, \apj, 143, 419
\bibitem[{Strigari et al.}, 2008]{Strigari08}Strigari, L.~E., Bullock, J.~S., Kaplinghat, M., Simon, J.~D., Geha, M., Willman, B., \& Walker, M.~G. 2008, \nat, 454, 1096 
\bibitem[{Sofue et al.}, 2004]{Sofue04}Sofue, Y., Kudoh, T., Kawamura, A., Shibata, K., \& Fujimoto, M. 2004, \pasj, 56, 633
\bibitem[{Wakker \& van Woerden}, 1997]{Wakker97}Wakker, B.~P., \& van Woerdan, H. 1997, \araa, 35, 217
\bibitem[{Wakker et al.}, 2008]{Wakker08}Wakker, B.~P., York, D.~G., Wilhelm, R., Barentine, J.~C., Richter, P., Beers, T.~C., Ivezi{\'c}, {\v{}Z}, \& Howk, J.~C. 2008, \apj, 672, 298
\bibitem[{Wolfire et al.}, 1995]{Wolfire95}Wolfire, M.~G., McKee, C.~F., Hollenbach, D., \& Tielens, A.~G.~G.~M. 1995, \apj, 453, 673
\bibitem[{Zait, Hoffman, \& Shlosman}, 2008]{Zait08}Zait, A., Hoffman, Y., \& Shlosman, I. 2008 \apj, 682, 835
\end{thebibliography}
\end{document}